# Strongly Confined Mid-infrared to Terahertz Phonon Polaritons in Ultra-thin $SrTiO_3$

Peiyi He[1,2#], Jiade Li[1,2#], Cong Li[3#], Ning Li[1,2], Bo Han[1,2], Ruochen Shi[1,2], Ruishi Qi[4], Jinlong Du[2], Pu Yu[3*], Peng Gao[1,2,5,6,7*]

[1] International Center for Quantum Materials, School of Physics, Peking University, Beijing 100871, China

[2] Electron Microscopy Laboratory, School of Physics, Peking University, Beijing 100871, China

[3] State Key Laboratory of Low-Dimensional Quantum Physics, Frontier Science Center for Quantum Information, and Department of Physics, Tsinghua University, Beijing 100871, China

[4] Department of Physics, University of California at Berkeley, Berkeley, CA 94720, USA

[5] Interdisciplinary Institute of Light-Element Quantum Materials and Research Center for Light-Element Advanced Materials, Peking University, Beijing 100871, China

[6] Collaborative Innovation Center of Quantum Matter, Beijing 100871, China

[7] Hefei National Laboratory, Hefei 230088, China

[#] These authors contributed equally to this work.

[*] Correspondence address: Pu Yu (yupu@mail.tsinghua.edu.cn), Peng Gao (pgao@pku.edu.cn)

## Abstract

Phonon polaritons (PhPs) enable subwavelength light control for infrared sensing, imaging, and optoelectronics, but conventional polar materials have narrow Reststrahlen bands, limiting applications. Materials that support PhPs with broad spectral range, strong field confinement, slow group velocity, and high quality factor are therefore needed. Here, using monochromatic electron energy loss spectroscopy in a scanning transmission electron microscope, we demonstrate that ultra-thin $SrTiO_3$ membranes possess the desired properties. Systematic measurements across varying thicknesses reveal two PhP branches with wide spectral dispersion, strong confinement, and anomalously slow group velocities spanning from the mid-infrared to terahertz range. Notably, in 3-nm-thick membranes, these polaritons exhibit unprecedented confinement factors exceeding 500 and group velocities as low as ~ $7 \times 10^{-5}c$, rivaling the best-performing van der Waals materials. These findings establish perovskite oxide such as $SrTiO_3$ as versatile platforms for tailoring light-matter interactions at the nanoscale, providing critical insights for the design of next-generation photonic devices requiring broadband operation and enhanced optical confinement.

## Teaser

Few-Unit-Cell freestanding perovskite membrane traps mid-IR and THz light with record-breaking confinement and group velocity.

## Introduction

Phonon polaritons (PhPs) are hybrid optical modes formed by the strong coupling between photons and optical phonons in polar materials. They exhibit strong light-matter interactions (*1*), subwavelength confinement (*2*), low-loss propagation (*3*), and tunable dispersion relations (*4*), with frequencies spanning the mid-infrared (MIR) to terahertz (THz) range. These unique properties enable a broad range of applications including optical imaging (*5*), sensing (*6-8*), data storage (*9*), and coherent thermal emission (*10*). Surface PhPs, as surface evanescent modes, typically arise within the Reststrahlen band (RB), a frequency range between transverse optical (TO) and longitudinal optical (LO) phonon modes, where the real part of permittivity is negative. However, for most polar materials, their RBs are relatively narrow and cannot meet the requirements for applications in a wide spectral range. Therefore, searching and developing novel materials with wide spectral ranges and high confinement capabilities have become critical for advancing PhP-based applications.

The $ABO_3$-type perovskite oxides may be able to break through the limitations of the narrow spectral range of PhPs. Firstly, the frequencies of polar vibrations of the octahedral oxygens and the central B-site metal atom of the perovskite oxides cover the MIR to THz range (*11*). Secondly, due to the giant LO-TO splitting caused by the large effective ionic charge, the PhPs of perovskite oxides span a wide RB range (*12*). These intrinsic properties of perovskite oxides are expected to result in a broad PhP spectral range spanning the MIR to THz region. Strontium titanate ($SrTiO_3$), as a prototypical perovskite oxide, has recently been predicted to be a promising PhP platform (*13*) in addition to its quantum paraelectricity (*14*), superconductivity (*15*), high electron mobility (*16*), *etc*. Notably, $SrTiO_3$ possesses two broad RBs extending from MIR to THz range, which allows for the simultaneous excitation of PhPs in two different frequency ranges as shown in Fig. 1A, opening up possibilities for MIR to THz applications. Intriguingly, with decreasing thickness, the PhPs of $SrTiO_3$ thin membranes are predicted to exhibit extraordinary properties such as high confinement, low group velocity, and excellent propagation quality (*13*), placing $SrTiO_3$ on a par with the two-dimensional van der Waals materials (*17, 18*). Note that polaritons with slow group velocity—i.e., a form of "slow light"—are a promising solution for time-domain processing of optical signals, such as all-optical tunable delays and optical buffers, while the increased light–matter interaction time also enables enhanced nonlinear effects and the miniaturization of integrated photonic components such as modulators and interferometers (*19-21*). These capabilities make the exploration of PhPs with ultra-slow group velocities particularly important for advancing infrared nanophotonic technologies.

However, a comprehensive understanding of the PhPs of $SrTiO_3$ and achievement of excellent properties face dual difficulties of not only the sample preparation but also experimental measurement.

A recent study reported the measurement of PhP dispersion in the MIR region from 70 to 74 meV for ~ 100 nm thick $SrTiO_3$ membranes (*22*). However, extraordinarily high confinement and slow group velocity of PhPs are only expected in the ultra-thin samples (*13*). Unlike van der Waals materials, $SrTiO_3$ is a three-dimensional ionic material, making it challenging to fabricate into nanosized thin sheets. This difficulty arises from its strong ionic bonds and three-dimensional lattice structure, which contrast with the weak interlayer van der Waals interactions in layered materials. Luckily, the recent advances in oxide epitaxy and transfer by using a soluble sacrificial layer (*23, 24*), have made this possible. The remaining difficulty lies in the direct detection of PhPs down to the THz spectral range. Due to the limitations of light sources and detectors, mainstream near-field optical methods have difficulty reducing frequencies below the MIR region, which is known as the "THz gap" (*25*). Although previous studies have investigated the MIR region of $SrTiO_3$ membranes (*22*), the THz region is still unexplored. The characteristics of PhPs in $SrTiO_3$ membranes within the THz range and their optical performance in ultra-thin samples remain largely unknown, which motivates the present study.

Electron energy loss spectroscopy incorporated in a scanning transmission electron microscope (STEM-EELS) has been proven to be a powerful tool for studying PhPs (*26-33*), enabling atomic-level spatial resolution and a larger momentum range. More importantly, since the PhPs detection by STEM-EELS relies on the continuous energy loss of inelastic scattering of incident electrons, it is not limited to the so-called "THz gap", which brings great advantages to the study of PhPs with a broad spectral range from the MIR to THz region.

Here, we extend the measurement capabilities of STEM-EELS to an energy threshold below 20 meV, enabling the simultaneous detection of two branches of PhPs in ultra-thin $SrTiO_3$ membranes, spanning from the MIR to THz range. Owing to the large momentum transfer provided by electron beam excitation, we observe broad dispersion ranges for both PhPs modes across membranes of varying thickness, nearly covering the entire RB ranges. Remarkably, these thin-film PhPs exhibit ultra-high confinement and ultra-slow group velocities. In membranes with the thickness of 3 nm (~ 8 unit cells), we observe PhPs with confinement factors exceeding 500 and group velocities as low as $7\times10^{-5}c$, which are the highest record so far to best of our knowledge. These findings demonstrate the potential of ultra-thin $SrTiO_3$ membranes for future applications in nanophotonics and advanced light manipulation technologies.

## Results

### PhPs in nano-thick $SrTiO_3$ membrane

We begin by discussing the PhP characteristics of $SrTiO_3$ based on its dielectric properties. Fig.

1B shows the real part and imaginary part of the dielectric functions of $SrTiO_3$. In the 12 ~ 59 meV ($TO_1$ to $LO_2$) and 68 ~ 99 meV ($TO_3$ to $LO_3$) frequency ranges (gray area, labeled as $RB_1$ and $RB_2$, respectively), the real part of the dielectric function is negative, allowing for the sustention of surface PhPs. The $RB_1$ region corresponds mainly to the vibrations of Sr and Ti atoms, while the $RB_2$ region is attributed to the vibrations of O atoms. Notably, $RB_1$ and $RB_2$ extend into the MIR and THz regions, with $RB_1$ having a broader spectral range than $RB_2$ due to very strong Sr-related TO mode (*34*). The dispersions of PhPs are confined within the $RB_1$ and $RB_2$ regions. This makes the THz band of $SrTiO_3$ particularly promising, as its accessible PhP range is comparable to, or even broader than, the MIR Reststrahlen bands of other state-of-the-art polaritonic materials such as h-BN (~ 25 meV) (*35*), $\alpha$-$MoO_3$ (~ 37 meV) (*33*), and SiC (~ 22 meV) (*32*). When the thickness of $SrTiO_3$ changes, the dielectric screening in the vertical direction is altered, which in turn modifies the dispersion of the PhPs. To further evaluate the dispersion behaviors of the PhPs in $SrTiO_3$, we calculated the analytical PhP dispersion of freestanding $SrTiO_3$ with various thicknesses by the complex momenta $q(\omega) + i\kappa(\omega)$ (see Methods). The real part $q(\omega)$ is shown in Fig. 1C. For membranes, the modes supported by the structure generally hybridize between the top and bottom surfaces and form symmetric and antisymmetric thin-film branches, with stronger splitting at smaller thicknesses. In Fig. 1C, the branches with negative group velocities (dashed lines) correspond to the symmetric modes, while the branches with positive group velocities (solid lines) correspond to the antisymmetric modes (see Supplementary Text 1 and Fig. S1 for more discussion). In the following, our analysis focuses on the antisymmetric branches, which are the only branches experimentally accessible in our measurement due to the large damping of the symmetric modes, as discussed in ref.(*36*).

We find that as the thickness decreases, the dispersion of thin-film PhPs becomes increasingly flatter. At the monolayer limit, the system will exhibit an ultra-slow group velocity (~ $10^{-5}c$, $c$ is the speed of light in the free space) comparable to that of monolayer h-BN (refs.(*13, 30, 37*)). In addition, our calculations show that the thin-film PhPs in the MIR region (MIR-PhPs) have a quality factor [$Q = q(\omega)/\kappa(\omega)$] of up to ~ 6 (Fig. 1D), which is close to experimentally observed values (2 ~ 6) (*22*). Intriguingly, we find that the $Q$ factor of PhPs in the THz region (THz-PhPs) can reach up to ~ 25, which is more than four times larger than that of MIR-PhPs. This suggests that the propagation quality of PhPs in $SrTiO_3$ slab holds promise to approach that of plasmons in h-BN-encapsulated graphene (~ 25) (*38*) and may even exceed that of the PhPs in h-BN (~ 20) (*39*) and $\alpha$-$MoO_3$ (~ 20) (*40*). The broad RB frequency ranges and excellent $Q$ factor of $SrTiO_3$'s PhPs highlight its tremendous potential, warranting further experimental investigation to fully explore its promising applications in nanophotonics.

**STEM-EELS measurements of thin-film PhPs**

To comprehensively investigate the thin-film PhPs in $SrTiO_3$ across THz and MIR regions, we conducted STEM-EELS measurements on freestanding nano-thick $SrTiO_3$ membranes. $SrTiO_3$ membranes were grown by pulsed laser deposition method and subsequently transferred onto lacy carbon TEM grids (see Methods). Figure 2A illustrates the setup of the STEM-EELS measurements. In our experiment, a monochromatic electron beam with an energy of 60 keV (energy resolution of 8 meV, see Supplementary Fig. 2A) was incident on the sample. After interaction with the sample, the high-angle annular dark-field (HAADF) images and EELS spectra were collected simultaneously. As a typical example, we first measured a freestanding $SrTiO_3$ membrane with a thickness of ~ 30 nm (see Fig. S2). As shown in Fig. 2B, the atomically resolved HAADF image near the sample boundary exhibits sharp edges and a wide-range flat surface, while the electron diffraction pattern (inset in Fig. 2B) also confirms the good crystallinity of the $SrTiO_3$ membrane. Figure 2C shows the energy distribution curves (EDCs) measured at different locations corresponding to the positions marked in Fig. 2B: aloof (blue curve), near the sample edge (red curve), and deep within the sample edge (yellow curve). The EDCs show distinct and rich loss peaks from 12 ~ 110 meV, reflecting the interesting spatial distribution of the PhPs and intrinsic phonons.

To fully identify these peaks and capture their spatial evolution, we performed spatial EELS mapping on the $SrTiO_3$ membrane with a step size of 10 nm. The EDC stack (merged every two lines, effectively 20 nm step size) and 2D mapping image in Figs. 2D and 2E clearly show the spatial variation of the different loss peaks, allowing us to precisely identify the origins of each peak. Notably, two prominent peaks with distinct spatial energy variations were observed inside the sample (the closer to the boundary, the higher the energy), undoubtedly corresponding to the THz-PhPs and MIR-PhPs (blue and red dashed lines in Fig. 2D). In Fig. 2E, we have marked other characteristic peaks. Firstly, for the MIR region (around $RB_2$), we observe a dispersionless peak at the highest energy (~ 99 meV) that only appears within the sample, which corresponds to the $LO_3$ mode shown in Fig. 1B. At slightly lower energy, there is a nondispersive onset near 93 meV, corresponding to the surface optical phonon ($SO_3$) energy where $\mathrm{Re}(\varepsilon) = -1$. This onset manifests as a broad hump that arises at the $SO_3$ energy and can also be observed in thinner samples (see Fig. S5). It originates from the convolution of a series of modes extending into the slab, as detailed and illustrated in Figs. S4. At lower energies, in addition to the MIR-PhPs propagating along the interior of the sample, there are also dipolar modes propagating along the sample edge (see Fig. S4). Among these, the low-$q$ modes near the $TO_3$ energy, with longer spatial extension, contribute most strongly. After convolution, these modes give rise to the edge-localized peak around the $TO_3$ energy in Fig. 2E. Peaks in the THz region (around $RB_1$) exhibit similar behaviors, where the energies of $LO_2$ and $SO_2$ are very close [ $\mathrm{Re}(\varepsilon)$ is steeper], leading to the convolution of $LO_2$ into the feature, which manifests as a nondispersive onset near 59 meV (see Fig. S4). At lower energies, we also observe a series of continuum excitations, corresponding to dipolar modes propagating along the sample edge, with particularly strong signals near the $TO_2$ mode (~28 meV). Additionally, we can also observe lower-energy peaks down to 17 meV, which include $LO_1$ and other PhPs, though these will not be discussed in detail here. Figure 2F presents the simulated EELS spectra by the boundary element method (BEM) with the same configuration as the experiment (Methods), which match the measured EELS spectra very well, further confirming the validity of our analysis. Our EELS measurements, with high spatial and energy resolution, high detection efficiency, and a broad spectral range, provide clear evidence of the PhPs in $SrTiO_3$ membranes spanning from the MIR to THz region.

**Ultra-confined thin-film PhPs**

Next, we extract the thin-film PhP dispersion through the energy-loss variation of electron beam probe in real space. Our electron beam can be regarded as a broadband excitation source. At a fixed position, only the modes satisfying the resonance condition are strongly excited and thus prominently appear in the EELS spectrum at the corresponding energy. Specifically, here the excited PhPs propagate to the boundary, reflect with an additional phase shift of $\pi/4$ (ref.(*41*)), and interfere with the original wave. The constructive interference condition is therefore given by: $2qd + \pi/4 = 2\pi$, which means that for each distance from the boundary $d$, only the mode with wavevector $q$ satisfying this relation is selectively enhanced and detected, and its corresponding energy is manifested in the EELS measurements. Then we can convert the experimentally measured spatial distribution of PhPs into dispersion relation in momentum space. The large momentum transfer of STEM-EELS, reaching up to ~ $10^6$ $cm^{-1}$, enables the observation of nearly the whole PhP dispersion in $SrTiO_3$ membrane (Fig. 3A). To investigate the thickness dependence of the thin-film PhP dispersion, we measured $SrTiO_3$ membranes with thicknesses down to 8 and 3 nm (Figs. 3B and 3C, see Fig. S5 for the real-space data). We find that as the membrane thickness decreases, the dispersion of the PhPs becomes progressively flatter. The corresponding results by calculated imaginary part of the complex Fresnel reflection coefficient $\mathrm{Im}\left[r_p(q,\omega)\right]$ (Methods) are presented in Figs. 3D, E, and F, and exhibit good agreement with the experimental data. These results suggest that reducing the thickness enables stronger confinement and decreases the group velocity of PhPs.

Accurate dispersion measurements enable us to extract and study the confinement and deceleration factors, two key indicators that characterize the PhP properties (*13*). The confinement factor measures the compression of the wavelength of light trapped in the PhPs and is defined by the ratio of the momentum of the PhPs to the wave vector of free light, $q/q_0$, where $q_0$ is the wave vector of free light corresponding to the PhPs energy. The deceleration factor quantifies the reduction in the speed of light trapped in the PhPs and is defined by the ratio of the group velocity of the PhPs to the speed of free light $c$, $D = v_g/c = c^{-1}\partial\omega/\partial q$, where $v_g$ and $\omega$ are the group velocity and the frequency of the PhPs. Figures 4A and 4B present the confinement and deceleration factors, respectively, of both THz-PhPs and MIR-PhPs in freestanding $SrTiO_3$ membranes with thicknesses of ~ 30 nm, ~ 8 nm, and ~ 3 nm. The experimental results (solid circles) exhibit good agreement with the calculated results (solid curves). For both THz-PhPs and MIR-PhPs modes, thinner samples show higher confinement factors and lower deceleration factors at a given energy. This underscores the superior capacity of thinner membranes to compress the light wavelength and slow down its propagation speed. Notably, for all sample thicknesses measured in this experiment, both THz-PhPs and MIR-PhPs readily achieve a confinement factor of 100, with the ~ 3 nm sample even surpassing a confinement factor of 500 (Fig. 4A). Moreover, the deceleration factors for both THz-PhPs and MIR-PhPs across all thicknesses can easily be reduced to the ~ $10^{-4}$ range (Fig. 4B), with the MIR-PhPs in the ~ 3 nm sample reaching as low as $7 \times 10^{-5}$ (inset of Fig. 4B). Such extremely low deceleration factors (group velocities ~ $10^{-5}\,c$) have only been reported in monolayer systems previously (*30*). It is important to note that our experiment is constrained by the edge quality of the samples (Fig. S6A), which limits practical access to higher-$q$ modes. Our reported confinement and group velocity reflect values that can be reliably achieved experimentally. In principle, better edge quality and additional measurements would enable even greater confinement and slower group velocity (see more discussion in Fig. S6). The PhPs in nano-thick freestanding $SrTiO_3$ membranes, spanning from the MIR to THz range, demonstrate remarkable ultra-high confinement and ultra-slow group velocity, which not only break the optical diffraction limit but also substantially enhance the light-matter interaction while reducing the propagation length of PhPs.

## Discussion

Recent optical measurements have reported results for MIR-PhPs in relatively thick (~ 100 nm) $SrTiO_3$ membranes (*22*). In our study, we have not only captured lower-energy THz-PhPs—difficult to observe optically due to the "THz gap"—but also revealed ultra-high confinement factors and ultra-low deceleration factors far surpassing those reported in previous optical studies (*22, 34, 42*). This success can be attributed to two key advances. First, we successfully achieved high-quality $SrTiO_3$ membranes with a thickness of only a few unit cells. Both previous reports and our findings have demonstrated that thinner samples exhibit flatter PhP dispersions, thereby enhancing their ability to compress the wavelength and slow down the light speed. Second, electron-beam excitation offers several advantages: high detection efficiency, excellent spatial resolution, large momentum compensation, and a broad spectral detection free from the "THz gap" limitation. These advantages enabled us to efficiently capture multi-mode polaritons and their dispersion across the full polariton band, revealing enhanced confinement and considerably slower propagation. Additionally, the high spatial resolution of energy-filtered EELS mapping allows us to observe how the localized PhP resonances in $SrTiO_3$ nanostructures can be effectively tuned through geometric structuring (see Fig. S7), providing insights for designing nanophotonic devices.

More importantly, the two branches of PhPs we observed in $SrTiO_3$ membranes exhibit distinct properties. Thanks to broad RBs of $SrTiO_3$, its PhP dispersion spans a wide range of 70 meV, covering from MIR to THz frequencies. This energy range is markedly larger than that reported in typical PhP systems, such as h-BN (*35*), $\alpha$-$MoO_3$ (*33*), SiC (*32*), and others (*28, 29*). Additionally, in ~ 3 nm thick $SrTiO_3$ membrane, we observed PhPs exhibiting ultra-high confinement factors surpassing 500 and ultra-slow group velocities down to ~ $7 \times 10^{-5}c$. These remarkable properties are on par with those of the prominent prototype h-BN(*30*). PhPs with such high confinement and slow group velocities can dramatically compress the wavelength and propagation length of light, facilitating the miniaturization of optical devices and enabling new quantum effects alongside strong light-matter interactions. Their silicon-compatible epitaxy further enhances technological relevance (*43*).

The tunability of PhPs is a key aspect for their widespread application in optical devices (*44-47*).

However, the intrinsic crystal properties of materials often pose great challenges in achieving effective tuning of these polaritons. In contrast, $SrTiO_3$, as a representative of perovskite oxides, can be easily tuned through various methods including doping (*48*), thermal (*49*), and electrostatic (*50*) techniques, substantially expanding the application potential of PhPs in perovskite oxide membranes. Our research highlights that perovskite oxide systems can serve as an emerging platform of PhPs, comparable to van der Waals materials (*17, 18*), opening up promising avenues for ultra-thin metasurfaces and enhanced light-matter interactions.

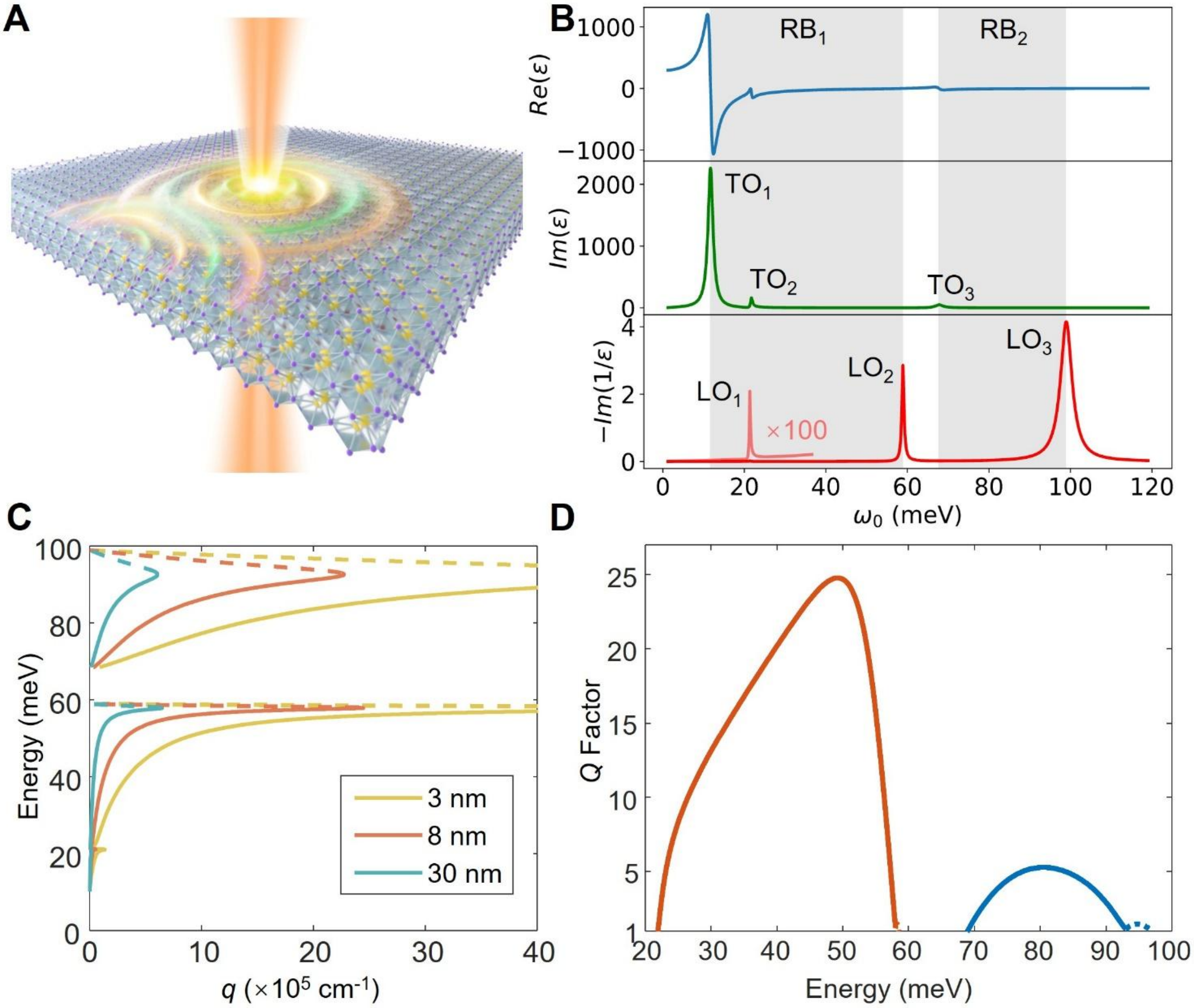

**Fig. 1. PhPs properties in freestanding $SrTiO_3$ membranes.** (**A**) Schematic illustration of PhPs excited in $SrTiO_3$ membrane. (**B**) Dielectric properties of $SrTiO_3$, where the upper, middle, and lower panels represent the real part of the dielectric function, the imaginary part of the dielectric function, and the loss function, respectively, with the optical phonon modes indicated beside. The gray shades correspond to the range of the RB. (**C**) Calculated dispersion of thin-film PhPs for freestanding $SrTiO_3$ with different thicknesses. (**D**) Calculated quality factor of thin-film PhPs in freestanding $SrTiO_3$. In (C) and (D), dashed lines denote symmetric modes and solid lines denote antisymmetric modes.

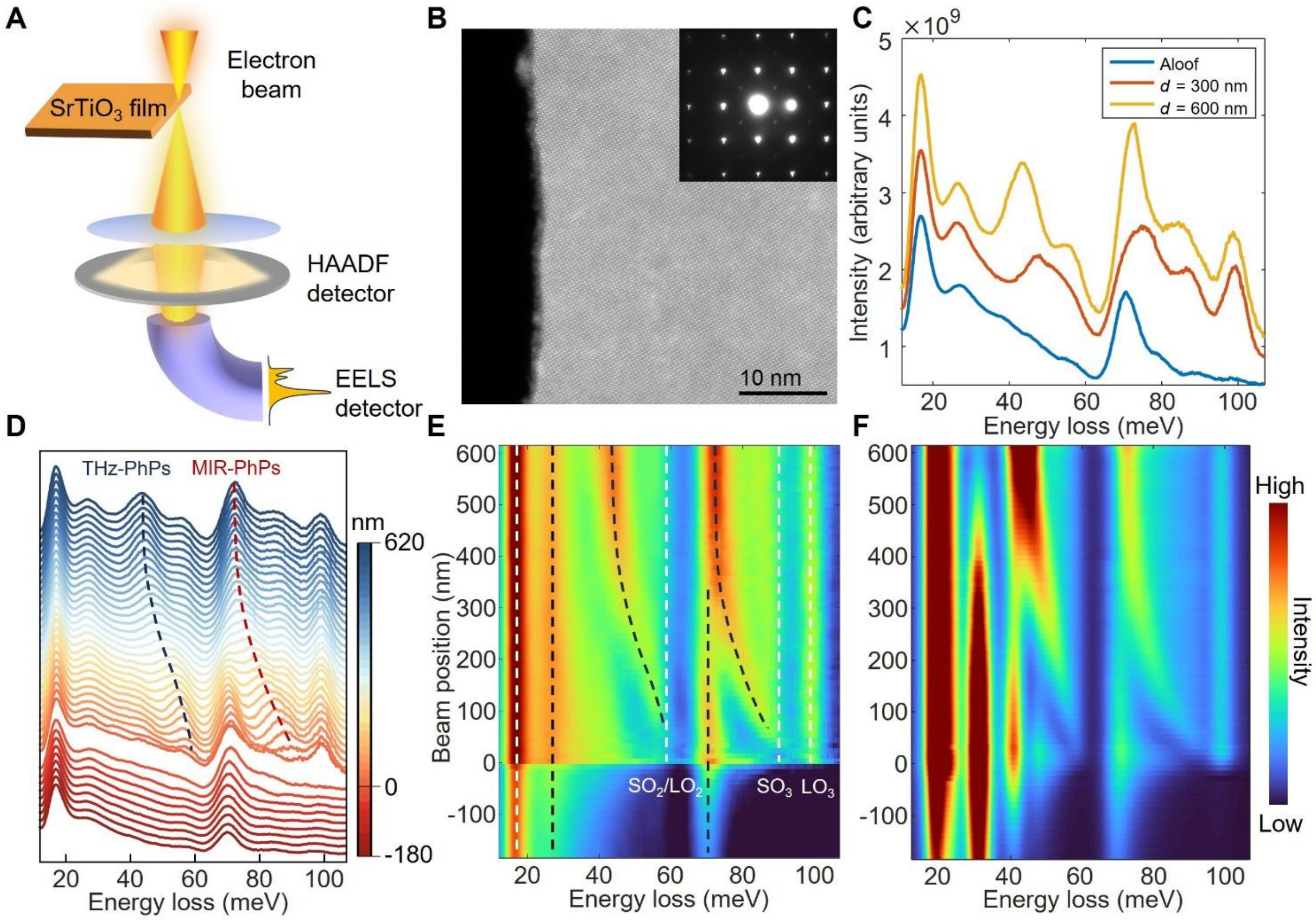

**Fig. 2. STEM-EELS measurements of thin-film PhPs in a 30 nm thick freestanding $SrTiO_3$ membrane.** (**A**) Schematic of the STEM-EELS setup, where the HAADF image and EELS spectrum of freestanding $SrTiO_3$ membrane can be simultaneously acquired. (**B**) Atomically resolved HAADF image near the edge of the $SrTiO_3$ membrane, with the inset showing the electron diffraction pattern. (**C**) EDC collected at different positions: blue represents the aloof configuration, while red and yellow correspond to spectra with the beam penetrating the sample at positions 300 nm and 600 nm from the edge, respectively. (**D**) EDC stack obtained from a vertical scan perpendicular to the edge, spanning from vacuum (180 nm from the edge) into the sample interior (up to 620 nm). The blue and red dashed lines serve as guides for the eye, indicating THz-PhPs and MIR-PhPs, respectively. An apparent gap near the edge (0 nm) arises because spectra acquired on the sample exhibit a higher residual background, resulting in an elevated baseline. (**E**) 2D visualization of the experimental dataset in (D), with the dashed lines serving as guides for the eye. (**F**) BEM simulation of the EELS probability under the same configuration as in (E).

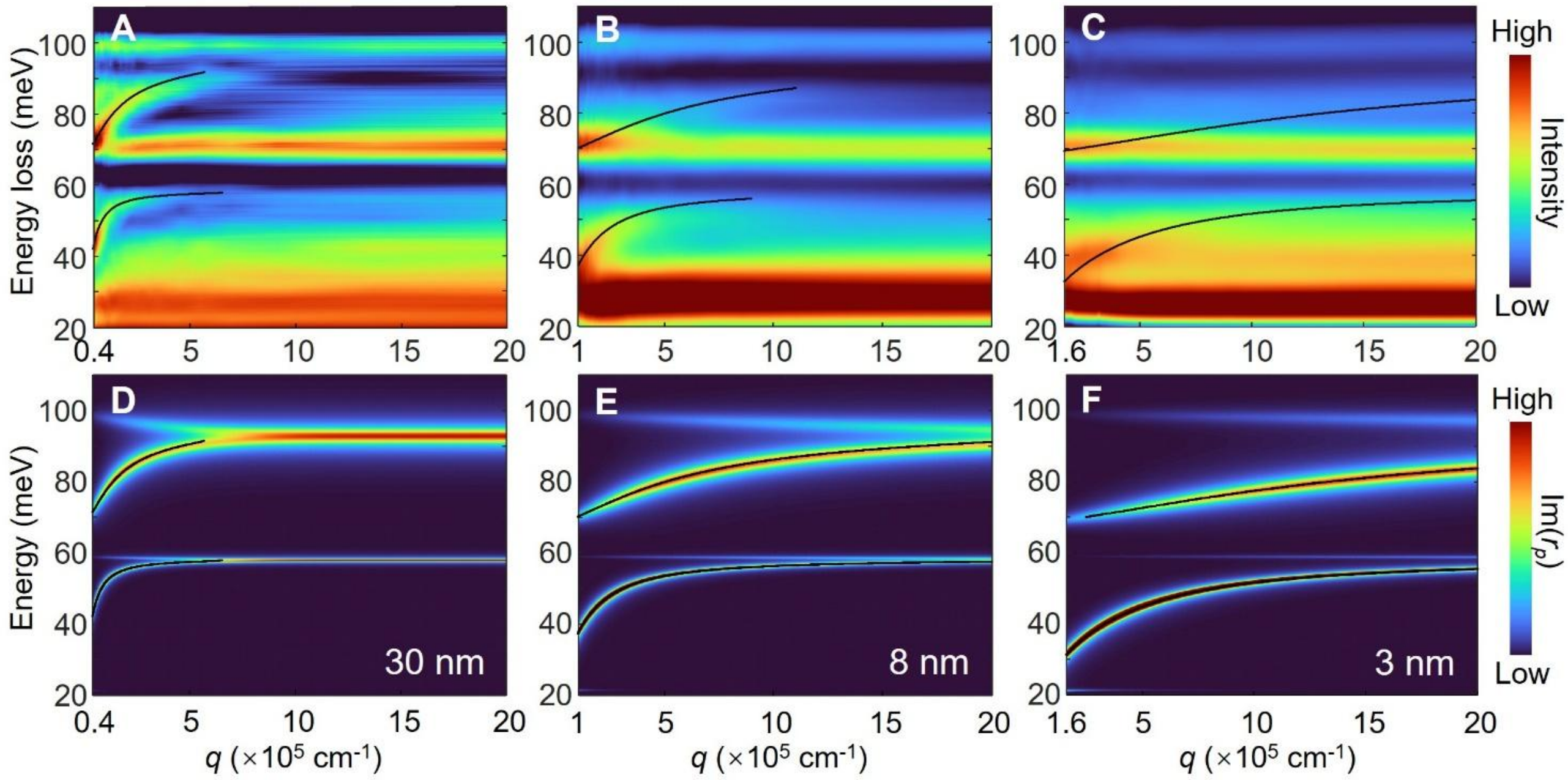

**Fig. 3. Thin-film PhP dispersion in freestanding $SrTiO_3$ membranes with different thicknesses.** (**A** to **C**) Dispersion relations of thin-film PhPs in $SrTiO_3$ with thicknesses of 30 nm, 8 nm, and 3 nm, respectively, transformed from the real-space experimental data. (**D** to **F**) Calculated imaginary part of the reflection coefficient, showing the dispersion relations of PhPs for membrane thicknesses corresponding to (A to C), respectively. The black lines represent analytically calculated dispersion curves.

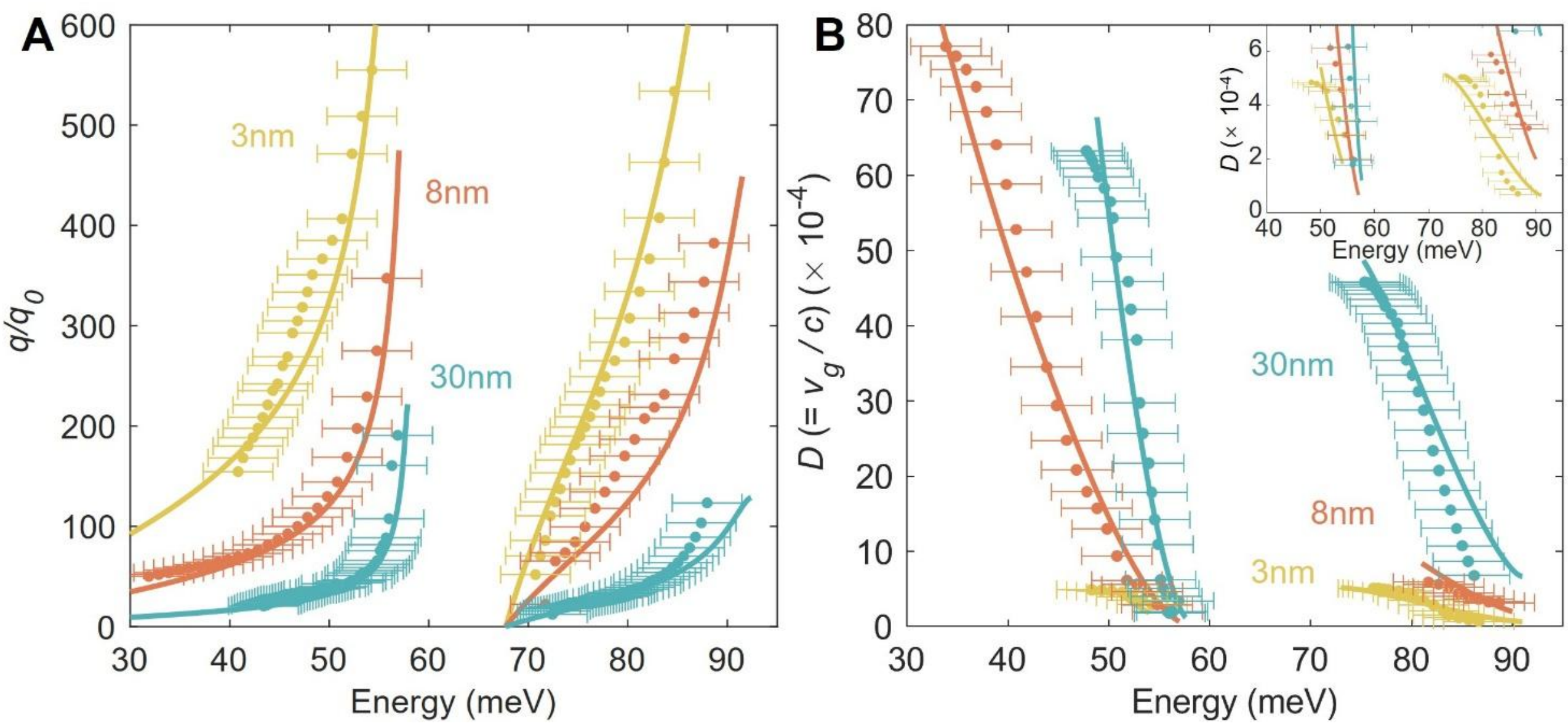

**Fig. 4. Confinement factor and deceleration factor of thin-film PhPs.** (**A**) Confinement factor of thin-film PhPs for different thicknesses. The dots with error bars represent experimental data, while the solid curves correspond to theoretical calculations. The error bars correspond to the energy resolution. (**B**) Deceleration factor of thin-film PhPs derived from dispersion in (A). The inset provides an enlarged view of the slow group velocity region.

## Materials and Methods

### The preparation of ultra-thin freestanding $SrTiO_3$ membranes for plan-view STEM.

To obtain ultra-thin $SrTiO_3$ freestanding membranes, heterostructures of $SrTiO_3$ (30 nm, 8 nm and 3 nm)/$SrCoO_{2.5}$ were grown on LSAT(001) substrates via the pulsed laser deposition method (ref.(*24*)). The $SrCoO_{2.5}$ layers were grown at 730 °C under an oxygen partial pressure of 13 Pa with a laser (KrF, $\lambda$ = 248 nm) energy density of 0.9 J cm$^{-2}$ and a repetition rate of 3 Hz. When finishing the growth of $SrCoO_{2.5}$ layers, the samples were gradually cooled to 700 °C with a cooling rate of 10 °C min$^{-1}$, and then the $SrTiO_3$ layers were deposited under the same condition as $SrCoO_{2.5}$ layer. After deposition, the samples were cooled to room temperature under the growth pressure condition with a cooling rate of 10 °C min$^{-1}$. To prepare the membranes from the heterostructures, the surface of the samples was faced down and attached to the carbon film side of a TEM grid. Then the sample and TEM grid were immersed together into the 36 % acetic acid at room temperature until the sacrificial layer was entirely dissolved. Afterward, freestanding $SrTiO_3$ membranes would be collected by the underneath carbon TEM grid (*23*).

### EELS and imaging experiments

We carried out STEM-EELS experiments on a Nion U-HERMES200 instrument operated at 60 kV. We employed a convergence semi-angle $\alpha = 20$ mrad and a collection semi-angle $\beta = 25$ mrad for all EELS datasets. For PhP measurements, the energy dispersion was set as 0.5 meV/channel and the typical energy resolution was about 8 meV. For thickness determination, the energy dispersion was 0.5 eV/channel. HAADF images were all acquired at the conditions of $\alpha = 35$ mrad and $\beta = (80, 210)$ mrad.

### EELS data processing

Custom-written MATLAB codes were used to process all acquired EELS datasets. EELS spectra were first aligned by cross-correlation and then normalized by the integrated intensity of the ZLP. Subsequently, the block-matching and 3D filtering (BM3D) algorithm was applied to remove Gaussian noise. After denoising, all vibrational spectra in the main text were obtained by multiplying the square of energy to better present the low-energy signals. Lucy-Richardson deconvolution was then employed to remove the broadening effect caused by the finite energy resolution. To verify the robustness of the processing, we also employed a background fitting and subtraction method to extract the signal, as shown in Fig. S3B. $\exp[P_3(x)]$ function was used to fit the background, with the selected fitting ranges of 19 ~ 24 meV and 122 ~ 137 meV, where $P_3(x)$ is a cubic polynomial with its coefficients as fitting parameters. Additionally, we applied spatial drift correction to obtain the line-scan results (i.e., summing the data along the y-direction). For thickness determination, we used the log-ratio method (*51*) and obtained the thickness from the EELS spectra in the energy range of -10 eV to 95 eV, processed with DigitalMicrograph software.

**Analytic calculation**

The $SrTiO_3$ permittivity can be approximated by (*51, 52*)

$$\varepsilon(\omega) = \varepsilon_\infty \times \prod_{i=1}^{3} \frac{\omega_{\mathrm{LO},i}^2 - \omega^2 - i\gamma_{\mathrm{LO},i}\omega}{\omega_{\mathrm{TO},i}^2 - \omega^2 - i\gamma_{\mathrm{TO},i}\omega}, \quad (1)$$

where $\varepsilon_\infty$, $\omega_{\mathrm{TO}}$, $\omega_{\mathrm{LO}}$, $\gamma_{\mathrm{TO}}$ and $\gamma_{\mathrm{LO}}$ denote the high-frequency permittivity, the energy of TO and LO phonon and the damping factor of TO and LO phonon respectively, and $\mathrm{i} = 1, 2, 3$ marks three pairs of TO and LO phonon. All these parameters are taken from ref.(*22*).

For a freestanding isotropic membrane of thickness $d$ in vacuum, by solving Maxwell's equations under the appropriate boundary conditions (*36, 53*) or by considering the photon momentum along the z-axis together with the Fabry-Perot quantization condition (*54*), we obtain the following expression for the polariton dispersion at large $q$ approximation:

$$q(\omega) + i\kappa(\omega) = \pm \frac{i}{d}\left[2 \arctan\left(\frac{i}{\varepsilon}\right) + \pi l\right], \quad (2)$$

where $l$ is an integer corresponding to the symmetric and antisymmetric modes. Specifically, $l = 0$ with the positive sign gives the antisymmetric mode, while $l = -1$ with the negative sign gives the symmetric mode. Each mode is supported within different energy ranges according to its respective decay behavior. A more detailed derivation and discussion can be found in Supplementary Text 1. Consequently, we have the analytical thin-film PhP dispersion for different thicknesses:

$$q(\omega) = \mathrm{Re}\left[-\frac{2}{d}\mathrm{arctanh}\left(\frac{1}{\varepsilon}\right)\right]. \quad (3)$$

The imaginary part of the complex Fresnel reflection coefficient $\mathrm{Im}\left[r_p(q,\omega)\right]$ can simultaneously display dispersion and damping on a color map. Using the Fresnel equations, we can derive the following for our freestanding slab (*54, 55*):

$$r_p = r\frac{1 - e^{i2k^z d}}{1 - r^2 e^{i2k^z d}}, \quad (4)$$

where

$$r = \frac{\varepsilon k_0^z - k^z}{\varepsilon k_0^z + k^z}$$

$$k_0^z = \sqrt{\left(\frac{\omega}{c}\right)^2 - q^2}$$

$$k^z = \sqrt{\varepsilon\left(\frac{\omega}{c}\right)^2 - q^2}$$

**BEM simulation**

We used the MNPBEM Toolbox in MATLAB to simulate EELS spectra by solving Maxwell's equations via BEM for isotropic dielectrics (*56*). The dielectric function used in the simulation was obtained from the results above. Approximately 5000 boundary elements were used for the calculation, resulting in EELS spectra at different incident positions. The data were then processed to obtain line-scan results and mode mapping results.

## Acknowledgments

This work was supported by the National Natural Science Foundation of China (52125307 to P.G.; 52025024 and 52388201 to P.Y.; 12504198 to J.L.) and the China Postdoctoral Science Foundation (Grant No. GZB20240028 and No. 2025M773365, J.L.). We acknowledge the High-Performance Computing Platform of Peking University for providing computational resources for part of the theoretical calculations. P.G. acknowledges the support from the Xplorer Prize.

## Author contributions

P.G. and P.Y. conceived the project. P.H. and J.L. performed STEM-EELS experiments assisted by N.L., B.H. and J.D. under the direction of P.G. C.L. grew the heterostructures and fabricated the $SrTiO_3$ membranes under the supervision of P.Y. P.H. and J.L. performed data processing and analysis assisted by R.S. and R.Q. P.H. performed MNPBEM simulations and analytic calculations under the guidance of P.G. P.H. and J.L. wrote the manuscript under the direction of P.G. All authors discussed the results at all stages and participated in the development of the manuscript.

## Competing interests

The authors declare no competing interests.

## Data and materials availability

All data needed to evaluate the conclusions in the paper are present in the paper and/or the Supplementary Materials.

# Supplementary Materials for

## Strongly Confined Mid-infrared to Terahertz Phonon Polaritons in Ultra-thin $SrTiO_3$

Peiyi He *et al.*

*Corresponding author: Pu Yu yupu@mail.tsinghua.edu.cn;
Peng Gao, pgao@pku.edu.cn

**This PDF file includes:**

**Supplementary Text**

1. Dispersion equation for small $d$

For an isotropic dielectric membrane of thickness $d$ suspended in air, the symmetric and antisymmetric modes are governed by the following relations (*36, 53*)

$$\varepsilon k_0^z + k^z \tanh\left(\frac{-ik^z d}{2}\right) = 0 \text{ (symmetric)}$$

$$\varepsilon k_0^z + k^z \coth\left(\frac{-ik^z d}{2}\right) = 0 \text{ (antisymmetric)},$$

$$k_0^z = \sqrt{\left(\frac{\omega}{c}\right)^2 - \tilde{q}^2},\ k^z = \sqrt{\varepsilon\left(\frac{\omega}{c}\right)^2 - \tilde{q}^2}, \tag{S1}$$

where $\varepsilon = \varepsilon(\omega)$ is the complex dielectric function of the membrane, and $\tilde{q} = q(\omega) + i\kappa(\omega)$ is the complex wavevector.

For very small $d$, the modes of interest have large wavevector. Under large wavevector approximation, we have $k_0^z \approx k^z \approx \pm i\tilde{q}$. In the following equations, we have chosen the signs to ensure decaying solutions such that $\mathrm{Im}(k^z)$, $\mathrm{Im}(k_0^z)$ and $\mathrm{Im}(\tilde{q})$ are all positive, which were further verified after obtaining the solutions. The dispersion equations then simplify to:

$$\tanh\left(\frac{-\tilde{q}d}{2}\right) = -\varepsilon \text{ (symmetric)}$$

$$\coth\left(\frac{\tilde{q}d}{2}\right) = -\varepsilon \text{ (antisymmetric)}. \tag{S2}$$

Using the transformation relations of hyperbolic functions and taking into account the multivalued nature of the solutions, we obtain the following expressions for the complex wavevector $\tilde{q}$

$$\tilde{q} = \frac{2}{d}\operatorname{arctanh}\left(\frac{1}{\varepsilon}\right) - \frac{1}{d}(2m-1)\pi \mathrm{i} \text{ (symmetric)}$$

$$\tilde{q} = -\frac{2}{d}\operatorname{arctanh}\left(\frac{1}{\varepsilon}\right) + \frac{1}{d}2n\pi \mathrm{i} \text{ (antisymmetric)}. \tag{S3}$$

Considering modes that do not vanish rapidly, we have $m = 0$ for symmetric mode and $n = 0$ for antisymmetric mode:

$$\tilde{q} = \frac{2}{d}\operatorname{arctanh}\left(\frac{1}{\varepsilon}\right) + \frac{1}{d}\pi \mathrm{i} \text{ (symmetric)}$$

$$\tilde{q} = -\frac{2}{d}\operatorname{arctanh}\left(\frac{1}{\varepsilon}\right) \text{ (antisymmetric)}. \tag{S4}$$

It is worth noting that in the small $d$ limit, the real parts of the symmetric and antisymmetric modes share the same functional form (set to be positive for plotting), reproducing the expression given in the main text (Eq. 3): $q(\omega) = \mathrm{Re}[-\frac{2}{d}\operatorname{arctanh}\left(\frac{1}{\varepsilon}\right)]$.

We can also obtain the expression for the imaginary part of the wavevector as

$$\kappa(\omega) = \mathrm{Im}\left[\frac{2}{d}\operatorname{arctanh}(\varepsilon)\right] \text{ (symmetric)}$$

$$\kappa(\omega) = \mathrm{Im}\left[-\frac{2}{d}\operatorname{arctanh}\left(\frac{1}{\varepsilon}\right)\right] \text{ (antisymmetric)}. \tag{S5}$$

These results can also be derived from the Fabry-Perot quantization condition (*54*). Here we have derived the solution presented in the main text, which clearly demonstrates how the thickness $d$ influences the PhP modes in the small-$d$ limit. For comparison, Fig. S1 also shows the solutions obtained by directly solving Eq. (S1) for different thicknesses, including the real part of wavevector $q(\omega)$, propagation length $L(\omega) = 1/(2\kappa(\omega))$, and factor $Q(\omega) = q(\omega)/\kappa(\omega)$. As the membrane thickness decreases, the $Q$ factor first drops and then becomes nearly independent of thickness. This behavior can be understood from the fact that, in the small-$d$ regime, both the real and imaginary parts of wavevector are squeezed by the $1/d$ scaling, so that their ratio no longer depends on $d$.

2. Mode mapping of localized PhP resonances

To further explore localized PhP resonances, we performed mode mapping on a freestanding $SrTiO_3$ membrane with a thickness of 30 nm. Two representative geometries were investigated: a protruding rectangular region with a side length of ~300 nm (Fig. S7A-C) and a triangular region with a side length of ~200 nm (Fig. S7D-F).

Since these protrusions are connected to the surrounding membrane, they do not constitute truly isolated blocks. Strictly speaking, the observed features are therefore not fully localized modes. Nevertheless, the energy-filtered EELS maps reveal distinct signatures that are consistent with our BEM simulations. For the rectangular geometry, the experimental maps (Fig. S7B) show that with increasing energy, the resonance hotspots gradually shift toward the corners, reminiscent of the dispersive nature of propagating PhPs (noting that the structure extends beyond the rectangular protrusion). For the triangular geometry, we observe a sequence of resonant features: at ~70 meV corresponding to a bulk-like mode, at ~75 meV showing strong localization at the corners, and at ~80 meV associated with edge resonances (Fig. S7E-F).

These results demonstrate that the high spatial resolution of STEM-EELS enables detailed mode mapping of PhPs, providing valuable insight into the behavior of confined resonances in nanostructured perovskite membranes. Such an approach is useful for understanding and optimizing the optical properties of nanostructured materials for advanced functionalities.

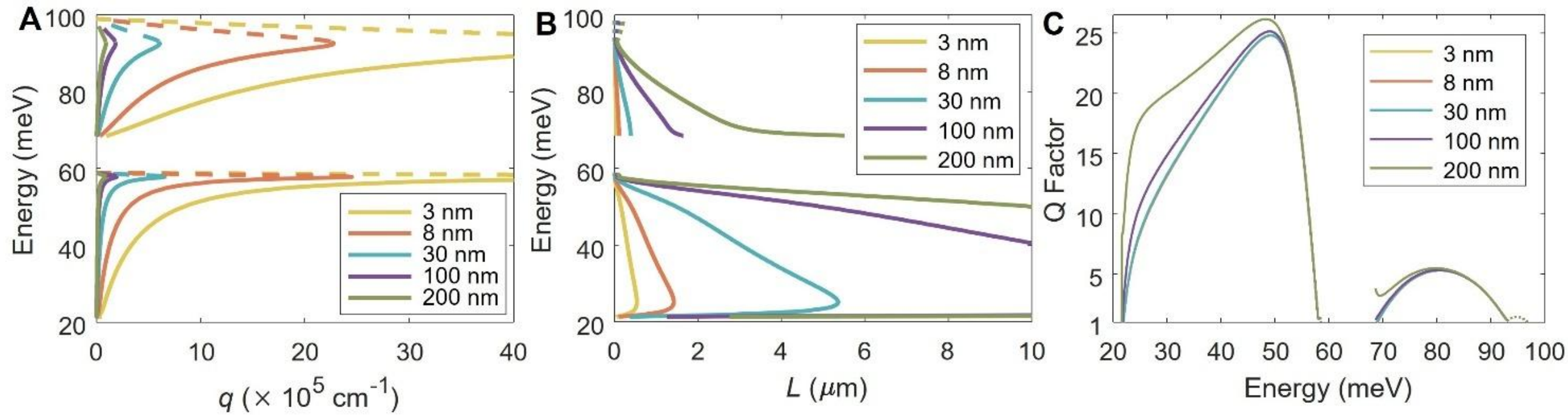

**Fig. S1. Thin-film PhPs in $SrTiO_3$ membranes.** Calculated (**A**) dispersion relations, (**B**) propagation lengths, and (**C**) quality factors of thin-film PhPs in freestanding $SrTiO_3$ membranes with different thicknesses. Dashed lines denote symmetric modes and solid lines denote antisymmetric modes.

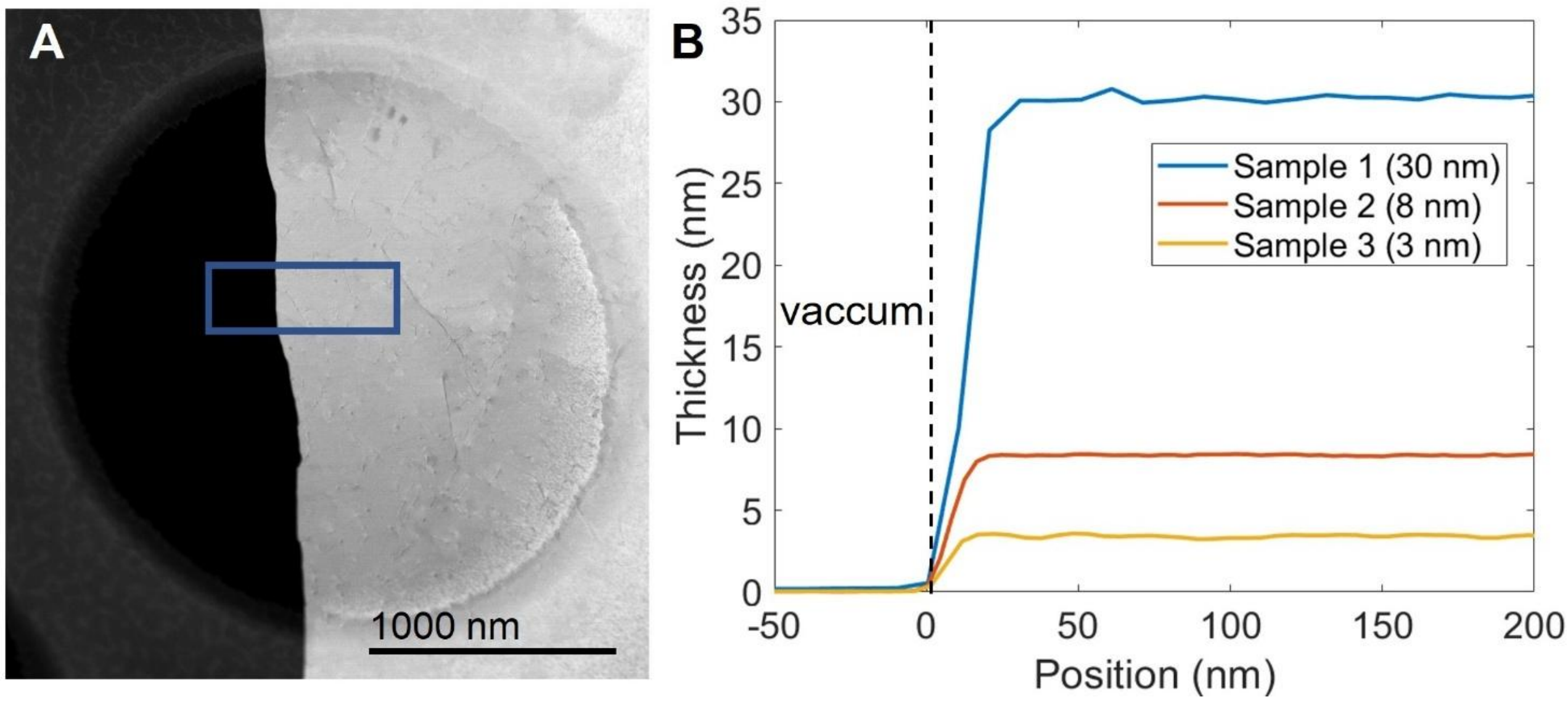

**Fig. S2. Sample configuration.** (**A**), HAADF image of a freestanding $SrTiO_3$ thin film, with part of the film resting on the carbon grid. The blue box indicates the STEM-EELS scanning region. (**B**), Thickness profiles of different samples determined by the log-ratio method, based on the ratio of plasmon losses to the ZLP intensity in EELS, obtained from scans spanning from vacuum to the sample interior.

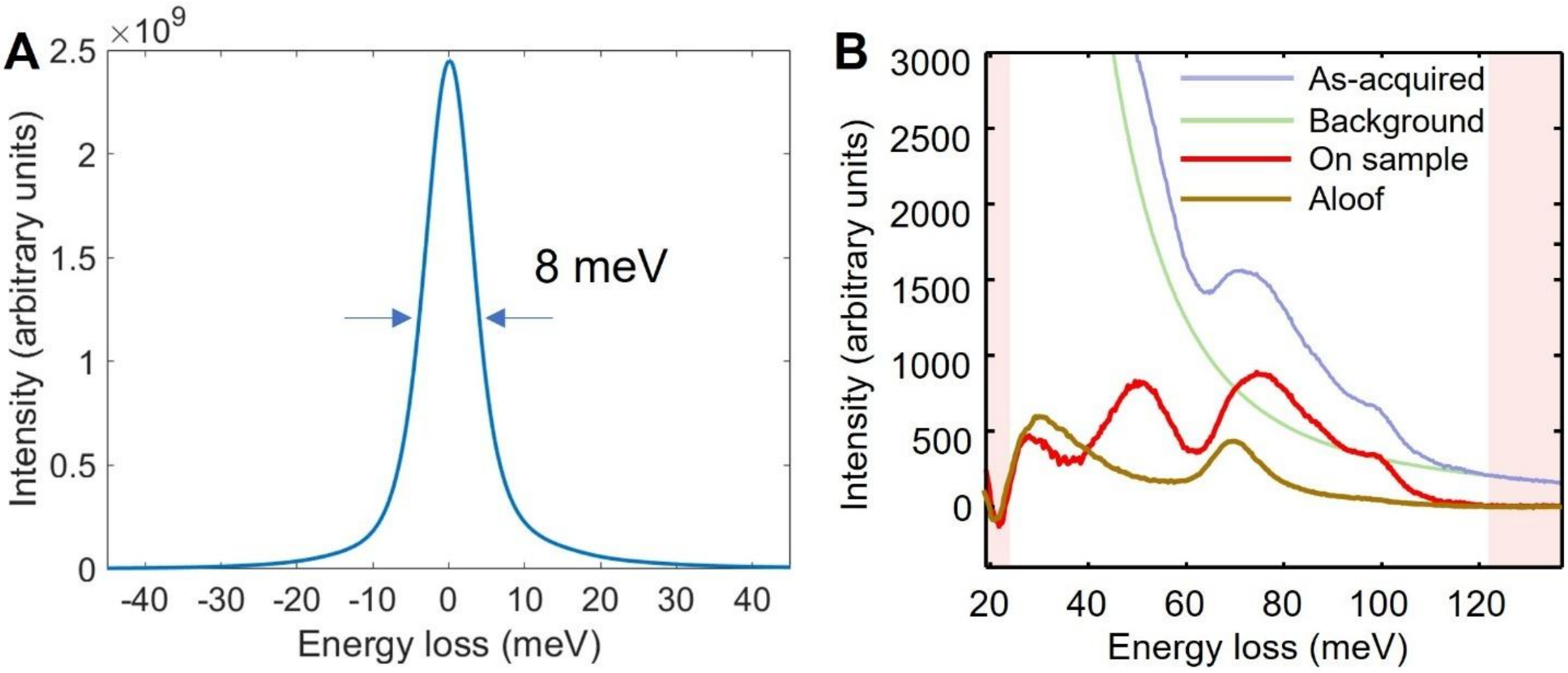

**Fig. S3. Energy resolution and background subtraction in STEM-EELS measurements.** (**A**), Measured zero-loss peak with a full width at half maximum of 8 meV, indicating the energy resolution of STEM-EELS measurements. (**B**), EELS spectra and the background subtraction. Purple: Raw data collected. Green: Fitted background using $\exp[P_3(x)]$ over two intervals. Red: Signal obtained after background subtraction. The above corresponds to the case where the electron beam passes through the sample. The brown curve represents the data obtained from the aloof configuration, processed in the same manner.

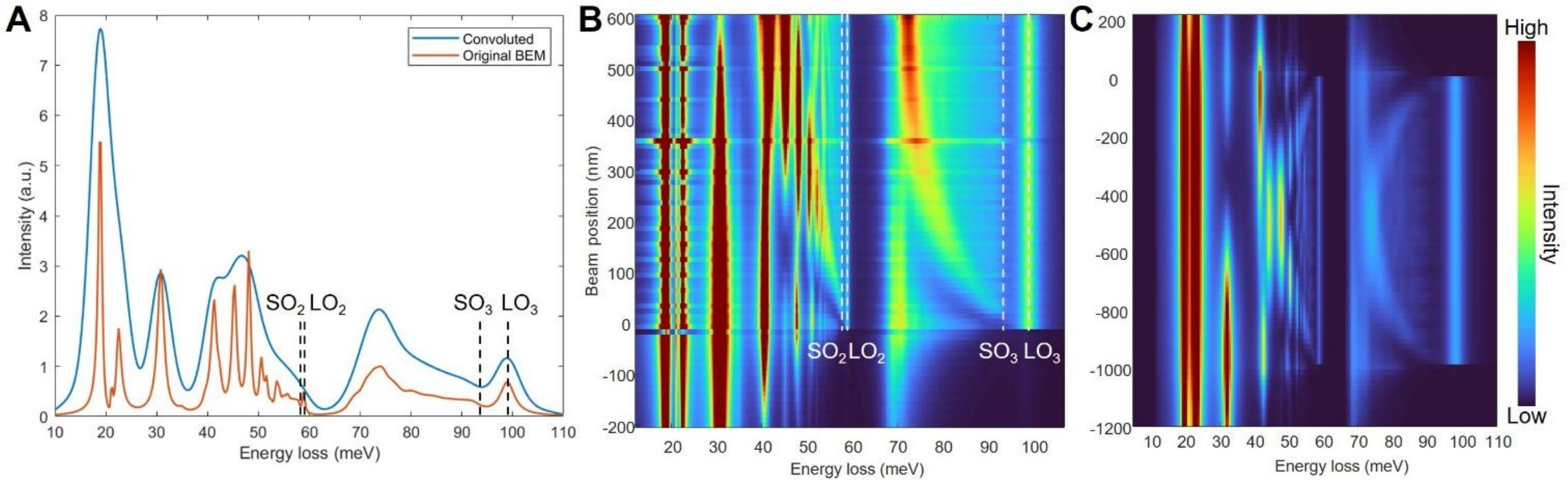

**Fig. S4. BEM simulations.** (**A**) Spectrum at 600 nm (from panel B) together with its convoluted counterpart. (**B**) Simulated EELS line scan corresponding to Fig. 2f before energy convolution. Modes satisfying $\boldsymbol{2qd+\pi/4=2n\pi}$ (with positive integers *n*) contribute to the feature in the figure. (**C**) Wide-range scan across a square plate of side length 1000 nm and thickness 30 nm. Due to the finite side length, access to lower-$q$, lower-energy modes is limited.

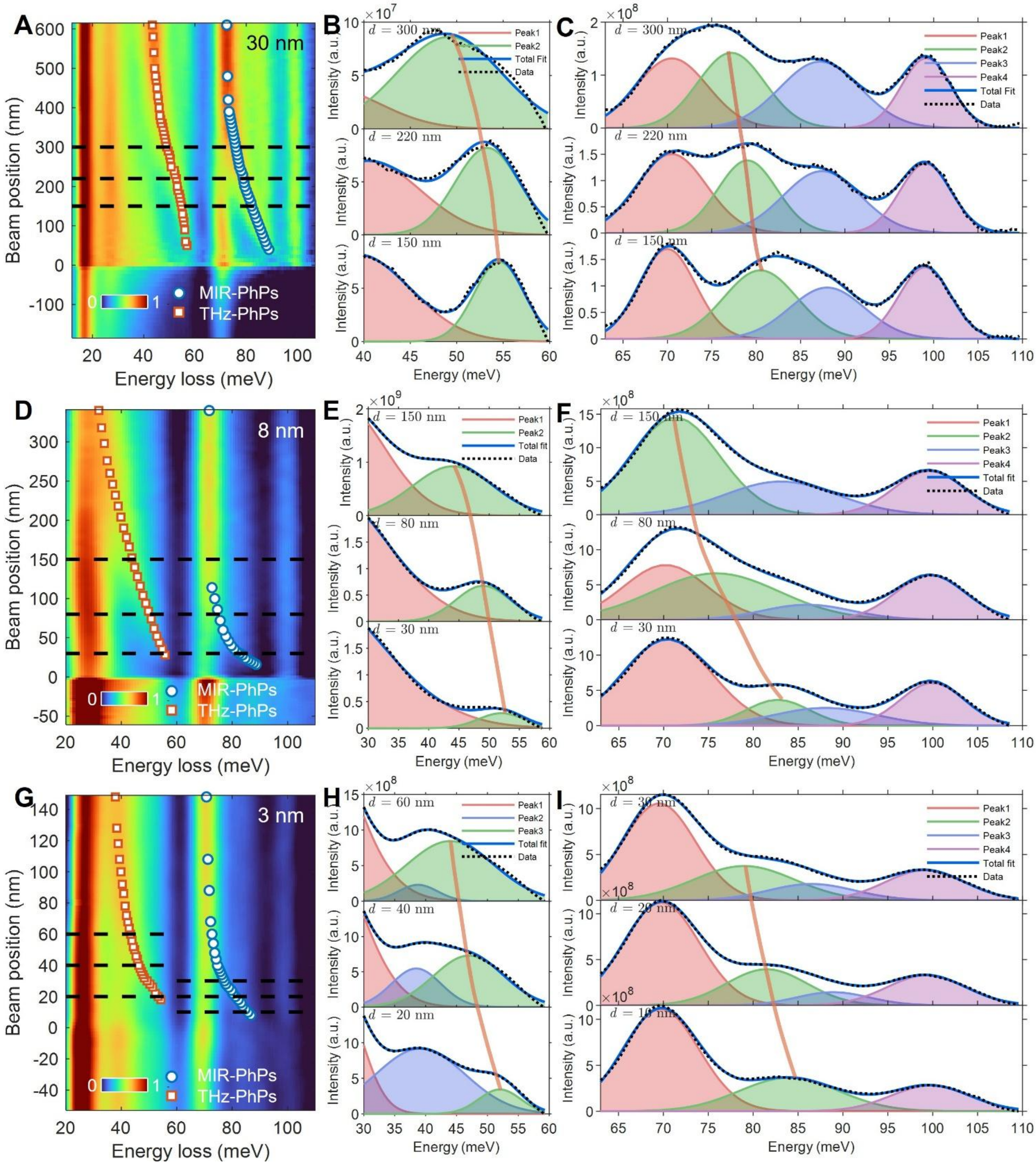

**Fig. S5. Data extraction and fitting for PhPs in $SrTiO_3$ membranes.** (**A** to **C**) 30 nm, (**D** to **F**) 8 nm, and (**G** to **I**) 3 nm membranes. (**A**, **D**, **G**) Line-scan EELS intensity maps as a function of beam position and energy loss. White circles denote MIR-PhP dispersions, and white squares denote THz-PhP dispersions, extracted from the subsequent fitting analysis. Note that for cases where the fits yielded the same energy, only the data point at the intermediate position was selected. Black dashed lines indicate the beam positions where representative spectra are analyzed. (**B**, **E**, **H**) Representative spectra in the THz range with multi-peak Gaussian fits. (**C**, **F**, **I**) Representative spectra in the MIR range with multi-peak Gaussian fits. Experimental data (black dotted lines) are decomposed into individual peaks (colored shaded areas), with the total fit shown as solid blue lines. The PhP-related

peaks are highlighted in green, and orange-red lines mark the shifts of their peak energies. The selected beam positions correspond to distances d from the boundary of (300, 220, 150) nm for 30 nm membranes, (150, 80, 30) nm for 8 nm membranes, and (60, 40, 20) / (30, 20, 10) nm for 3 nm membranes.

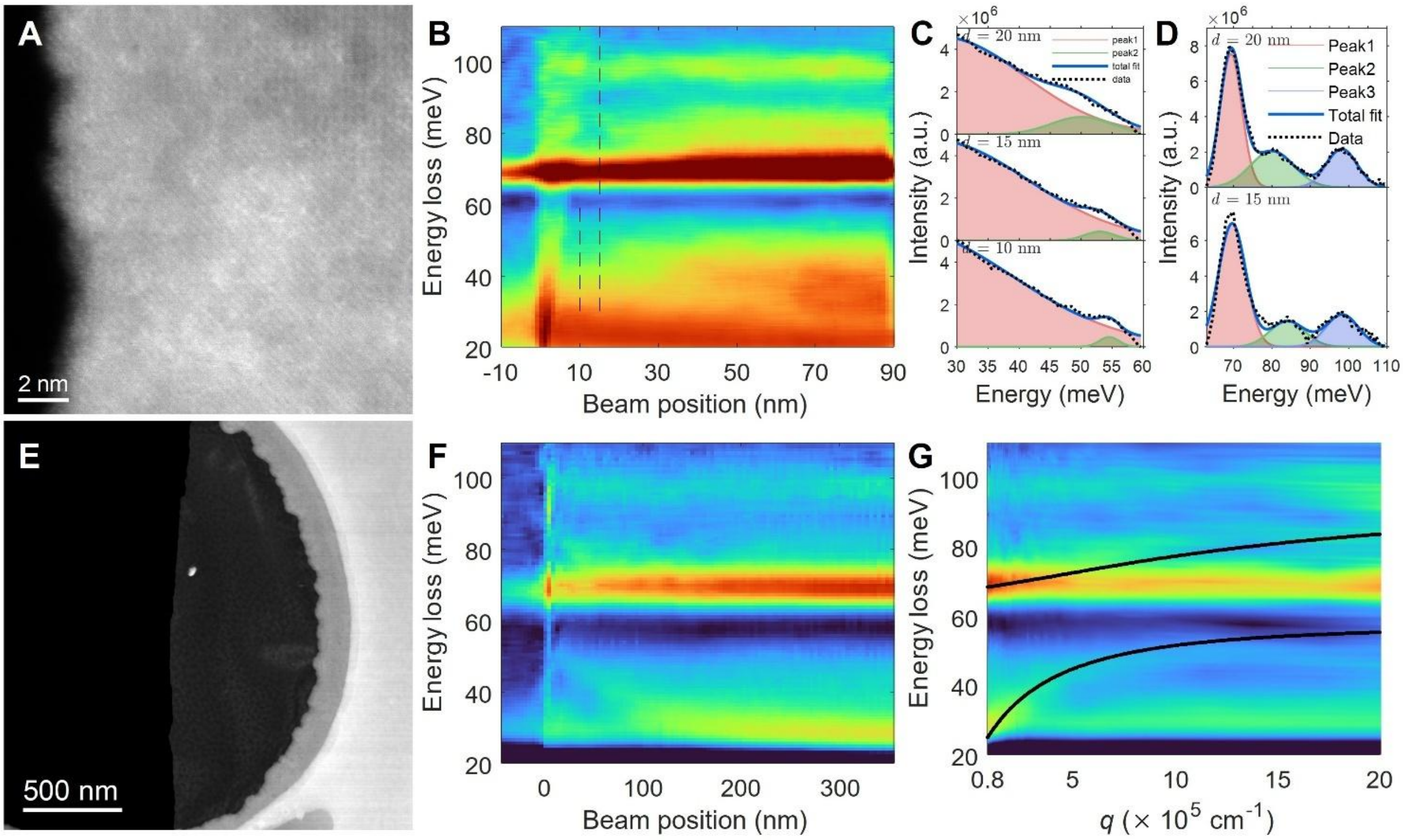

**Fig. S6. Additional datasets for 3-nm $SrTiO_3$ membranes showing the effects of spatial sampling and field of view on PhP measurements.** (**A**) STEM-HAADF image at the edge of the 3-nm membrane, revealing ~2-nm roughness. (**B**) EELS line-scan acquired with 1 nm per pixel across the edge. PhP features in the THz and MIR bands are visible. Black dashed lines mark the spectra plotted in (C to D). (**C**) Representative THz-band spectra with multi-peak Gaussian fits. (**D**) Representative MIR-band spectra with multi-peak Gaussian fits. (**E**) Overview STEM-HAADF image of the 3-nm membrane on a carbon grid. (**F**) Wide-range EELS line-scan (4 nm per pixel) over an extended field of view, enabling access to lower PhP energy. (**G**) PhP dispersion obtained from (F). Black curves are analytically calculated dispersion relations. These additional datasets were acquired with an updated direct electron detector and an optimized optical path, achieving an improved energy resolution of 5.6 meV. Edge roughness (~2 nm) limits the achievable confinement by introducing edge-position uncertainty (real-space to $q$ mapping), perturbing modes and increasing high-$q$ damping, and reducing fringe visibility/contrast. Random edge conditions make quantitative analysis more difficult. The reported confinement and group velocity values are conservative and represent reliably achievable levels.

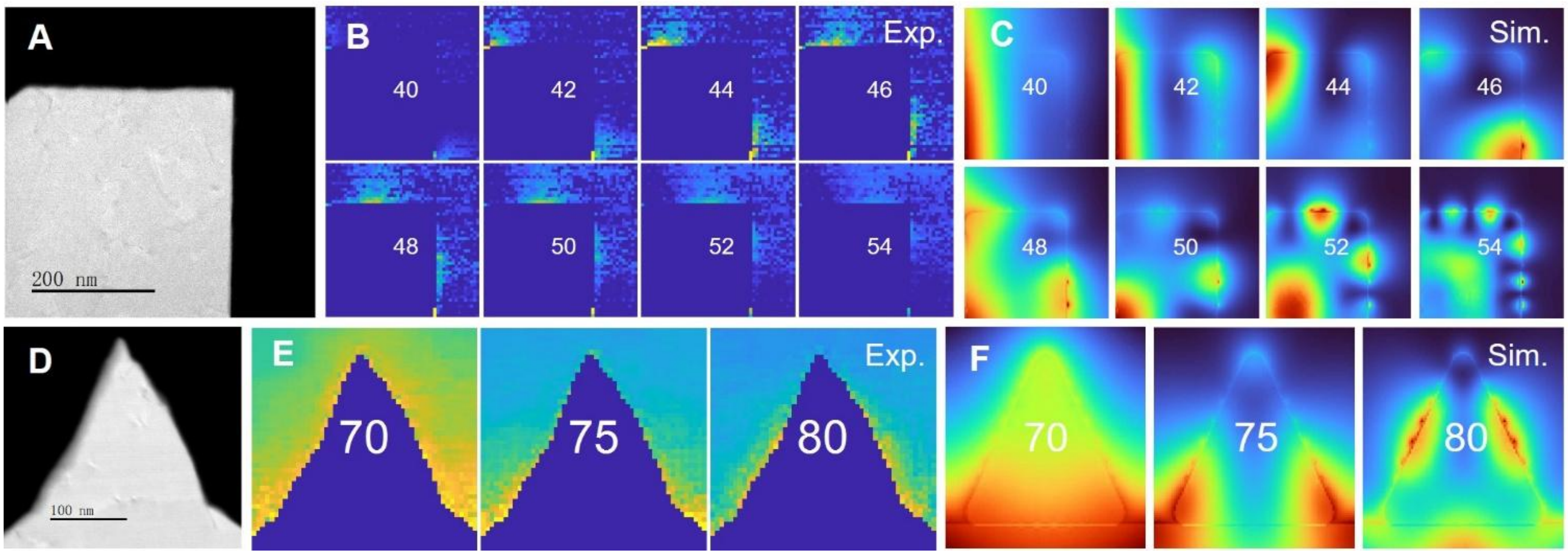

**Fig. S7. Real-space mode mapping on a 30 nm thick $SrTiO_3$ film.** (**A**), HAADF image of a protruding rectangle. (**B**), Energy-filtered EELS maps, with numbers representing the energy in meV (same for subsequent cases). (**C**), BEM simulation results corresponding to (B). (**D**), HAADF image of a protruding triangle. (**E**), Energy-filtered EELS maps. (**F**), BEM simulation results corresponding to (E).